\documentclass[aps,prd,preprint,nofootinbib]{revtex4}
\usepackage{amsfonts}
\usepackage{mathrsfs}
\usepackage{graphicx}
\usepackage{amsmath}
\usepackage{amssymb}
\usepackage{subfigure}
\usepackage{epsfig}
\usepackage{graphicx}
\newcommand{\bqa}{\begin{eqnarray}}
\newcommand{\eqa}{\end{eqnarray}}
\newcommand{\beq}{\begin{equation}}
\newcommand{\eeq}{\end{equation}}
\allowdisplaybreaks[1]
\graphicspath{{fig/}{dia/}} \DeclareGraphicsExtensions{.eps}
\begin{document}
\baselineskip 20pt
\title{NLO QCD corrections to exclusive electroproduction of quarkonium\\ [0.7cm]}

\author{Zi-Qiang Chen$^1$\footnote{chenziqiang13@mails.ucas.ac.cn} and Cong-Feng Qiao$^{1,2}$\footnote{qiaocf@ucas.ac.cn, corresponding author}}
\affiliation{$^1$ School of Physics, University of Chinese Academy of Science, Yuquan Road 19A, Beijing 10049 \\
$^2$ CAS Key Laboratory of Vacuum Physics, Beijing 100049, China}
\author{~\\}

\begin{abstract}
~\\ [-0.3cm]
The process of exclusive electroproduction of vector quarkonium (EEQ), $e p\to epV$, is per se an interesting topic in studies of quarkonium production mechanism, QCD description of diffractive interaction and nucleon structure. We investigate this process in the framework of nonrelativistic QCD and QCD collinear factorization at the next-to-leading order QCD accuracy.
The perturbative convergence behavior is discussed in a large range of photon virtuality $Q^2$.
The $J/\psi$ large-$Q^2$ electroproduction data at HERA can be well explained, and the $\Upsilon$ differential production rate is predicted. The uncertainties in theoretical predictions with radiative corrections are greatly reduced. Notice the EEQ process is extremely sensitive to the gluon distribution in nucleon, the generalized parton distribution, our results will constraint the gluon density with high precision while confronting to the future experimental data. For the sake of comparing convenience, the analytic expressions are provided.

\vspace {7mm} \noindent {PACS numbers: 13.60.-r, 14.20.Dh, 14.40.Pq}

\end{abstract}
\maketitle

\section{Introduction}
The processes of exclusive electro- and photo-production of quarkonium $\gamma^{(*)} p\to V p$ with $V=J/\psi$ or $\Upsilon$, are particular interesting and important but yet not well explored, especially the former. The photon here can be highly virtual for electroproduction or real for photoproduction. These processes provide unique opportunities in studying the quarkonium production mechanism, and perturbative QCD calculation reliability. They are experimentally adjustable in physical parameters, say for example the virtuality of the photon. Furthermore, they are gluon rich, hence extremely sensitive to the gluon distribution in the nucleon, particular to the off-diagonal effects. Stable theoretical calculations are necessary therefore to reduce the uncertainty of gluon density distribution in the still vague domain of small Bjorken variable $x_B$.

In experiment, $J/\psi$ production processes have been extensively studied at HERA, whereas for $\Upsilon$ production the data are limited to the photoproduction case. For a review, see for example \cite{exreview}. For future, some projects on deep inelastic experiment are in progress or proposed, like ENC at FAIR \cite{ENC}, eRHIC at BNL \cite{eRHIC}, LHeC at CERN \cite{LHeC} and EIC in China \cite{EicC}, where the EEQ process will be further explored attentively. Theoretically, two main frameworks are employed in the evaluation, that is the QCD collinear factorization \cite{cfs1} and BFKL $k_T$ factorization \cite{cch1,ce1}. Although the BFKL approach has a solid perturbative QCD foundation, can sum up large logarithms of energy ln(1/x) and implies the $k_T$ information of gluon in the nucleon in the description of hard diffractive processes \cite{ryskin,ktfrthr1,ktfrthr2,ktfrthr3}, it is impaired by the absence of full next-to-leading order (NLO) calculation, which however tends to be tough and may yield enormous corrections. Higher order calculation in collinear factorization hopefully can explain the existing $\Upsilon$ exclusive photoproduction data, but is fraught with difficulties for the $J/\psi$ case \cite{photoNLO1,photoNLO2,photoNLO3}. Whereas the exclusive quarkonium electroproduction processes, which in some sense are even more important in physics due to the adjustable virtuality of the intermediate photon, have not been explored properly. In this paper, we calculate the NLO QCD corrections to exclusive quarkonium electroproduction processes, and investigate their implications to the parton distribution in the nucleon.

According to QCD factorization, the EEQ processes may be allocated into three domains, where the purtabative calculable sector and nonperturbative part belonging to different energy scales are properly separated. Namely,

1) The hard partonic process $\gamma^* g(q)\to Q\bar{Q}g(q)$. Due to the hard scales provided by the heavy quark mass or by the photon virtuality $Q^2$, this part can be described by perturbative QCD (pQCD).

2) The transition from the $Q\bar{Q}$ pair to the physical quarkonium state. Herein, the transition probability can be encoded into the non-relativistic QCD (NRQCD) matrix element $\langle O^V\rangle$ \cite{NRQCD}.

3) The parton distribution within the nucleon. This effect may be separated from the hard process via $k_T$ factorization or collinear factorization mechanism.

It is worth mentioning that the parton distribution in the nucleon here refers to the so-called generalized parton distribution (GPD), as in the case of deeply virtual Compton scattering (DVCS) \cite{DVCS1,Jispin,DVCS3}. The GPD extends the usual forward PDF to the non-forward situation, and encode more richer information about the nucleon, like the nucleon spin \cite{Jispin}. The study of GPD is nowadays a very dynamical field, for reviews see for instance references \cite{review1,review2,review3}.

\section{Kinematics and factorization}

\begin{figure}
\hskip +1.5cm
\begin{center}
\includegraphics[scale=0.46]{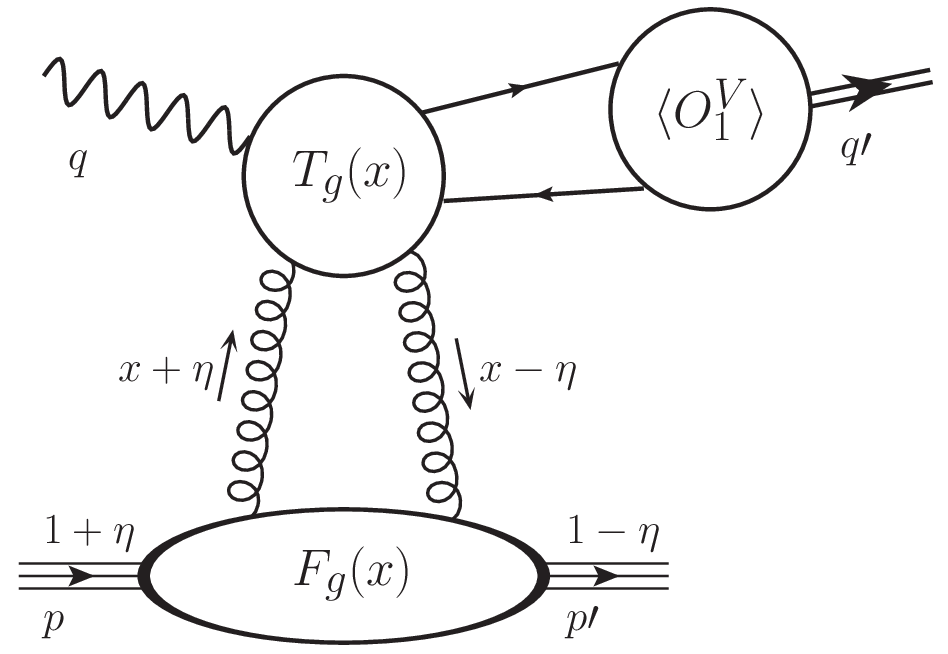}
\caption{Kinematics of quarkonium exclusive production. Momentum fractions $x$ and $\eta$ refer to $(\bar{p}\cdot n_-)$.}
\label{factorization}
\end{center}
\end{figure}

The kinematics of exclusive quarkonium production is schematically shown in Figure \ref{factorization}. The momenta of incident  photon and proton, outgoing quarkonium and proton, are denoted by $q$, $p$, $q^\prime$ and $p^\prime$ respectively. In the calculation, notations $\bar{q}=(q+q^\prime)/2$, $\bar{p}=(p+p^\prime)/2$ and $\Delta=p^\prime-p$ are also employed, and the following Lorentz invariants
\begin{align}
Q^2=&-q^2,\quad M^2=q^{\prime2},\quad m_N=p^2=p^{\prime2},\nonumber \\
&s=(q+p)^2=W^2,\quad t=\Delta^2
\end{align}
are then appeared in the analytic expression, where $M$ and $m_N$ are the mass of quarkonium and proton respectively.

Noticing that working in light-cone coordinate is convenient, we choose a frame where $\bar{q}$ and $\bar{p}$ to be collinear. By introducing two light like vectors $n_+=(1,0,0,1)\Lambda$ and $n_-=(1,0,0,-1)/(2\Lambda)$, where the arbitrary value parameter $\Lambda$ has the dimension of mass,  their momenta can be expanded as
\begin{eqnarray}
\bar{p}^\mu=n_+^\mu+\dfrac{m_N^2-t/4}{2}n_-^\mu\ ,\ \
\bar{q}^\mu=-\xi n_+^\mu+\dfrac{Q^2-M^2+t/2}{4\xi}n_-^\mu\ ,
\end{eqnarray}
and hence
\begin{eqnarray}
\Delta^\mu=-2\eta n_+^\mu+\eta(m_N^2-t/4)n_-^\mu+\Delta_{\perp}^\mu\ .
\end{eqnarray}
Here,
\begin{eqnarray}
\xi=-\frac{\bar{q}\cdot n_-}{\bar{p}\cdot n_-}\approx\frac{Q^2-M^2}{2s+Q^2-M^2}\ ,\ \
\eta=-\frac{1}{2}\frac{\Delta\cdot n_-}{\bar{p}\cdot n_-}\approx\frac{Q^2+M^2}{2s+Q^2-M^2}\ \ .
\end{eqnarray}
The skewedness parameter $\eta$ here plays a similar role as the Bjorken variable $x_B$ in deep-inelastic scattering. Following common usage, we use the term ``small $x_B$'' instead of ``small $\eta$'' in this paper.

According to the NRQCD and collinear factorization, the EEQ process amplitude can be expressed as
\begin{align}
\mathcal{M}^{\lambda\lambda^\prime}=&\dfrac{4\pi\alpha_s\sqrt{4\pi\alpha}e_q}{m}\sqrt{\dfrac{\langle O^V\rangle}{m}} \nonumber \\
&\quad\quad\quad\times\sum_{p=g,q}\int^1_{-1}dx(T_p^{\lambda\lambda^\prime}(x,\eta,\xi)F^p(x,\eta,t)
+\tilde{T}_p^{\lambda\lambda^\prime}(x,\eta,\xi)\tilde{F}^p(x,\eta,t))\ ,
\label{fac}
\end{align}
where $m$ is the mass of heavy quark equal to $M/2$ in the leading order of relativistic expansion, $e_q$ is heavy quark electric charge.
The superscript $\lambda$ ($\lambda^\prime$) denotes the helicity of incoming photon (outgoing quarkonium).
$T_p^{\lambda\lambda^\prime}$ ($\tilde{T}_p^{\lambda\lambda^\prime}$) and $F^p$ ($\tilde{F}^p$) represent the hard partonic amplitude and the matrix element of light-cone parton operators respectively. Their dependence on the factorization scale $\mu_F$ is suppressed for brevity. Following the convention and definition of Ref.\cite{review1}, $F^p$ and $\tilde{F}^p$ may be expressed as
\begin{align}
&F^p=\frac{1}{2(\bar{p}\cdot n_-)}\left[H^p(x,\eta,t)\bar{u}(p^\prime)n\!\!\!/_-u(p)+E^p(x,\eta,t)\bar{u}(p^\prime)\frac{i\sigma^{n_- \Delta}}{2m_N}u(p)\right],\nonumber \\
&\tilde{F}^p=\frac{1}{2(\bar{p}\cdot n_-)}\left[\tilde{H}^p(x,\eta,t)\bar{u}(p^\prime)n\!\!\!/_-\gamma_5u(p)+\tilde{E}^p(x,\eta,t)\bar{u}(p^\prime)\frac{\gamma_5(\Delta\cdot n_-)}{2m_N}u(p)\right],
\end{align}
where $H^p(x,\eta,t)$, $E^p(x,\eta,t)$, $\tilde{H}^p(x,\eta,t)$ and $\tilde{E}^p(x,\eta,t)$ are twist-2 GPDs with
arguments $x$ and $\eta$. The parton momentum fractions are represented by the combination of $x$ and $\eta$ as shown in  Fig.\ref{factorization}. The effect of higher-twist GPDs are generally complicated and small in comparison with the NLO perturbative corrections \cite{MJP}, which hence will not be taken account in this work.

Note, the symmetric properties of GPDs may simplify the calculation. The gluon distributions satisfy
\begin{equation}
H^g(x,\eta,t)=H^g(-x,\eta,t)\ ,\ \ \tilde{H}^g(x,\eta,t)=-\tilde{H}^g(-x,\eta,t)\ ,
\end{equation}
and similarly are the $E^g$ and $\tilde{E}^g$. For quark distributions, different combinations are considered usually in the calculation. That is
\begin{align}
&H^{q(+)}(x,\eta,t)=H^q(x,\eta,t)-H^q(-x,\eta,t)\ , \\
&\tilde{H}^{q(+)}(x,\eta,t)=\tilde{H}^q(x,\eta,t)+\tilde{H}^q(-x,\eta,t)\
\end{align}
referring to the ``singlet'' combination, and
\begin{align}
&H^{q(-)}(x,\eta,t)=H^q(x,\eta,t)+H^q(-x,\eta,t)\ , \\
&\tilde{H}^{q(-)}(x,\eta,t)=\tilde{H}^q(x,\eta,t)-\tilde{H}^q(-x,\eta,t)
\end{align}
for the ``nonsinglet'' case. As well, similar combinations exist for $E^q$ and $\tilde{E}^q$. Since photon and vector quarkonium have the same $C$ parities, only the $C$ even (singlet) component of quarks in GPD contributes to (\ref{fac}).

\section{Analytical results}

\begin{figure}
\centering
\subfigure[]{
\includegraphics[width=0.3\textwidth]{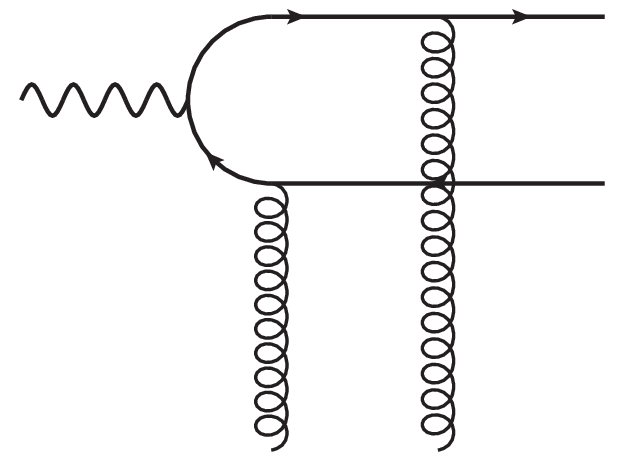}}
\subfigure[]{
\includegraphics[width=0.3\textwidth]{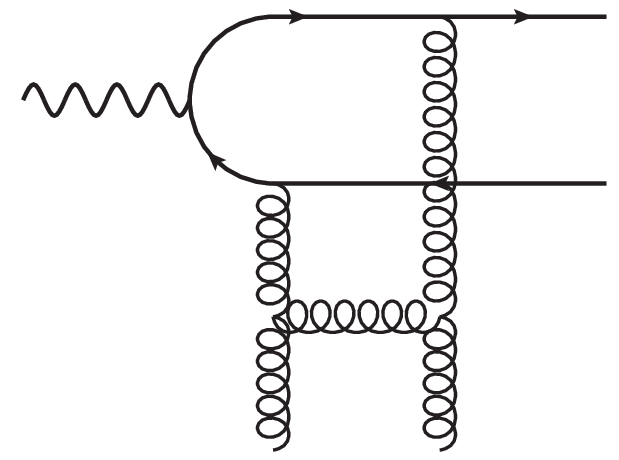}}
\subfigure[]{
\includegraphics[width=0.3\textwidth]{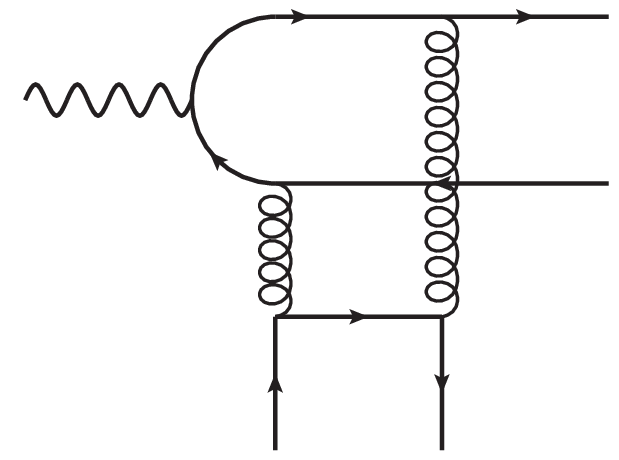}}
\caption{Typical Feynman diagrams for LO (a) and NLO (b, c) partonic processes.}
\label{feynman}
\end{figure}

The typical leading order (LO) and NLO Feynman diagrams are shown in Fig.\ref{feynman}. At LO, only gluon involved process contributes, whereas the light quark induced process may appear at NLO. The effects of intrinsic heavy quark inside the nucleon are reasonably small and will be neglected.

The convolution integration in (\ref{fac}) stretches from $|x|>\eta$ (DGLAP) region to the $|x|<\eta$ (ERBL) region. In $|x|<\eta$ region, the amplitude does not contain imaginary part, and the $i\varepsilon$ prescription in propagators can be dropped. The analytic continuation is performed by restoring the $i\varepsilon$ via
\begin{align}
&x\to x+i\varepsilon, \quad \text{for} \quad x>\eta\ , \nonumber \\
&x\to x-i\varepsilon, \quad \text{for} \quad x<-\eta\ .
\end{align}

The manipulation of calculation may be simplified greatly by making a coordinate transformation, namely
\begin{equation}
y_1=\frac{\eta-x}{\eta-\xi}\ ,\ \quad y_2=\frac{\eta+x}{\eta-\xi}\ .
\end{equation}
We then have
\begin{eqnarray}
&&T_g^{\lambda\lambda^\prime}(x,\eta,\xi)=\frac{1}{x^2-\eta^2} \left(A_g^{(0)\lambda\lambda^\prime}(y_1,y_2)+\frac{\alpha_s(\mu_R)} {\pi}A_g^{(1)\lambda\lambda^\prime}(y_1,y_2)\right)\ , \\
&&\tilde{T}^{\lambda\lambda^\prime}_g(x,\eta,\xi)=\frac{1}{x^2-\eta^2} \frac{\alpha_s(\mu_R)}{\pi}\tilde{A}_g^{(1)\lambda\lambda^\prime}(y_1,y_2)\ , \\
&&T_q^{\lambda\lambda^\prime}(x,\eta,\xi)=\frac{1}{\eta-\xi}
\frac{\alpha_s(\mu_R)} {\pi}A_q^{(1)\lambda\lambda^\prime}(y_1,y_2)\ ,\\
&&\tilde{T}^{\lambda\lambda^\prime}_q(x,\eta,\xi)=\frac{1}{\eta-\xi} \frac{\alpha_s(\mu_R)}{\pi}\tilde{A}_q^{(1)\lambda\lambda^\prime}(y_1,y_2)\ .
\end{eqnarray}
Here, $\mu_R$ is the renormalization scale. The helicity amplitudes obey $A_p^{++}=A_p^{--}$, $\tilde{A}_p^{++}=-\tilde{A}_p^{--}$ and $A^{0+}=A^{0-}=A^{+0}=A^{-0}=\tilde{A}^{0+}=\tilde{A}^{0-}=\tilde{A}^{+0}=\tilde{A}^{-0}=0$, which in accordance with the requirement of helicity conservation. There exist also crossing symmetries under the variable exchange $y_1\leftrightarrow y_2$ (i.e. $x\leftrightarrow -x$), i.e.,
$A_g(y_1,y_2) = A_g(y_2,y_1)$, $\tilde{A}_g(y_1,y_2) = -\tilde{A}_g(y_2,y_1)$, $A_q(y_1,y_2)=-A_q(y_2,y_1)$, $\tilde{A}_q(y_1,y_2)=\tilde{A}_q(y_2,y_1)$.

The LO result is pretty simple,
\begin{equation}
A_g^{(0)++}=A_g^{(0)--}=\frac{-1}{\sqrt{y_1+y_2-1}}A_g^{(0)00}=-\frac{1}{3}\frac{1}{y_1+y_2}\ .
\end{equation}

In the computation of NLO corrections, the ultraviolet (UV) and infrared (IR) singularities are regulated in
dimensional regularization with $D=4-2\epsilon$ prescription adopted. The following singular terms arising from one-loop diagrams contain both UV and IR singularities:
\begin{align}
A_{g,\text{loop-pole}}^{(1),++}&=A_{g,\text{loop-pole}}^{(1),--}=\frac{1}{\epsilon}\frac{C_A^2C_F}{24y_2(y_1+y_2)^2}\bigg((y_1^2+y_2^2)\text{ln}y_1\nonumber \\
&-\left(y_1^2+\frac{y_1y_2}{2}+\frac{y_2^2}{2}\right)\text{ln}(y_1+y_2)+\frac{3C_F}{8C_A}(y_1^2-3y_1y_2+2y_2)\bigg)+\{y_1\leftrightarrow y_2\}\ , \nonumber \\
A_{g,\text{loop-pole}}^{(1),00}&=-\frac{\sqrt{y_1+y_2-1}}{\epsilon}\frac{C_A^2C_F}{24y_2(y_1+y_2)^2}\bigg((y_1^2+y_2^2)\text{ln}y_1-(y_1^2+y_1y_2)\text{ln}(y_1+y_2)\nonumber \\
&-\frac{3C_F}{4C_A}(2y_1y_2-y_2)\bigg)+\{y_1\leftrightarrow y_2\}\ , \nonumber \\
A_{q,\text{loop-pole}}^{(1),++}&=A_{q,\text{loop-pole}}^{(1),--}=-\frac{A_{q,\text{loop-pole}}^{(1),00}}{\sqrt{y_1+y_2-1}} \nonumber \\
&=-\frac{1}{\epsilon}\frac{C_AC_F(y_1-y_2)}{9y_2(y_1+y_2)^2}\bigg(\text{ln}y_1-\frac{y_1+y_2}{2y_1}\text{ln}(y_1+y_2)\bigg)-\{y_1\leftrightarrow y_2\}\ ,\nonumber \\
\tilde{A}_{g,\text{loop-pole}}^{(1),++}&=-\tilde{A}_{g,\text{loop-pole}}^{(1),--}=-\frac{1}{\epsilon}\frac{C_AC_F^2}{64y_1}-\{y_1\leftrightarrow y_2\}\ ,\nonumber \\
\tilde{A}_{g,\text{loop-pole}}^{(1),00}&=\tilde{A}_{q,\text{loop-pole}}^{(1),00}=\tilde{A}_{q,\text{loop-pole}}^{(1),++}=\tilde{A}_{q,\text{loop-pole}}^{(1),--}=0\ .
\label{looppole}
\end{align}

The UV singularities are removed by renormalization. For the renormalization of heavy quark field, heavy quark mass and gluon field, we take the on-shell (OS) scheme; for the renormalization of coupling constant, the modified minimal-subtraction ($\overline{\text{MS}}$) scheme is used. The singular terms arising from counter terms are
\begin{align}
A_{g,\text{ct-pole}}^{(1),++}&=A_{g,\text{ct-pole}}^{(1),--}=-\frac{1}{\epsilon}\frac{C_AC_F^2}{32y_2(y_1+y_2)^2}\bigg(\frac{y_1^2}{2}-\frac{3y_1y_2}{2}+y_2-\frac{2\beta_0y_1y_2}{3C_F}\bigg)+\{y_1\leftrightarrow y_2\}\ , \nonumber \\
A_{g,\text{ct-pole}}^{(1),00}&=-\frac{\sqrt{y_1+y_2-1}}{\epsilon}\frac{C_AC_F^2}{16(y_1+y_2)^2}\bigg(y_1-\frac{1}{2}+\frac{\beta_0y_1}{3C_F}\bigg)+\{y_1\leftrightarrow y_2\}\ , \nonumber \\
\tilde{A}_{g,\text{ct-pole}}^{(1),++}&=-\tilde{A}_{g,\text{ct-pole}}^{(1),--}=-\tilde{A}_{g,\text{loop-pole}}^{(1),++}\ , \nonumber \\
A_{q,\text{ct-pole}}^{(1),++}&=A_{q,\text{ct-pole}}^{(1),--}=A_{q,\text{ct-pole}}^{(1),00}=\tilde{A}_{g,\text{ct-pole}}^{(1),00}=\tilde{A}_{q,\text{ct-pole}}^{(1),++}=\tilde{A}_{q,\text{ct-pole}}^{(1),--}=\tilde{A}_{q,\text{ct-pole}}^{(1),00}=0.
\label{ctpole}
\end{align}

After the cancellation between (\ref{looppole}) and (\ref{ctpole}), the remaining singularities will be absorbed into the parton distribution functions, achieved by introducing the scale dependent GPD, i.e.
\begin{equation}
F^p(x,\eta,\mu_F)=F^p(x,\eta)-\frac{1}{\epsilon}\left[\frac{\alpha_s}{2\pi}\frac{\Gamma(1-\epsilon)}{\Gamma(1-2\epsilon)}\left(\frac{4\pi\mu_R^2}{\mu_F^2}\right)
^\epsilon\right]\sum_{p^\prime}\int_{-1}^1dyV_{pp^\prime}(x,y,\eta)F^{p^\prime}(y,\eta)\ ,
\end{equation}
where $V_{pp^\prime}$ denotes the GPD evolution kernel.

Finally, the NLO result is finite and can be written in a general form
\begin{equation}
A^{(1)\lambda\lambda\prime}=\bigg(c_0^{\lambda\lambda\prime}+ \sum_{i=1}^{11}c_i^{\lambda\lambda\prime}f_i\bigg)\pm\{y_1\leftrightarrow y_2\}\ ,
\label{gen}
\end{equation}
where the plus sign corresponds to $A_g$ and $\tilde{A}_q$, while the minus to $A_q$ and $\tilde{A}_g$.
The coefficients $c_0$ and $c_i$ in (\ref{gen}) are a bit lengthy and are given in Appendix. The $f_i$ are either logrithm or polylogarithm functions yielded from the Feynman integration:
\begin{align}
&f_1=\sqrt{\tfrac{y_1-1}{y_1}}\text{ln}(\sqrt{y_1-1}+\sqrt{y_1}),\quad f_2=f_1^2,\nonumber\\
&f_3=\sqrt{\tfrac{y_1+y_2}{y_1+y_2-1}}\text{ln}(\sqrt{y_1+y_2-1}+\sqrt{y_1+y_2}),\quad f_4=f_3^2\ , \nonumber \\
&f_5=\text{ln}(\tfrac{y_1}{y_1+y_2}),\quad f_6=\text{ln}2+\text{ln}(y_1+y_2),\quad f_7=\text{ln}^2(y_1+y_2)\ ,\nonumber \\
&f_8=\text{ln}(\tfrac{4m^2}{\mu_F^2})\text{ln}(\tfrac{y_1}{y_1+y_2})+\text{ln}^2y_1,\quad f_9=\text{Li}_2(1-2y_1),\quad f_{10}=\text{Li}_2(1-2y_1-2y_2)\ ,\nonumber \\
&f_{11}=C_0(4(1-y_1-y_2),1-2y_1,1-2y_2,1,1,0)\ ,
\end{align}
with
\begin{align}
&C_0(4(1-y_1-y_2),1-2y_1,1-2y_2,1,1,0) = \nonumber \\
&\tfrac{1}{2(y_2-y_1)} \bigg\{\text{Li}_2\left[\tfrac{2y_1}{y_1+y_2}\right]
- \text{Li}_2\left[\tfrac{(y_1+y_2)(1-2y_2)} {y_1+y_2-4y_1y_2}\right]
+ \text{Li}_2\left[\tfrac{2y_1(1-2y_2)} {y_1+y_2-4y_1y_2}\right]\nonumber \\
& - \text{Li}_2\left[\tfrac{2y_1}{y_1+y_2 - \sqrt{(y_1+y_2)/(y_1+y_2-1)}(y_2-y_1)}\right]
- \text{Li}_2\left[\tfrac{2y_1}{y_1+y_2 + \sqrt{(y_1+y_2)/(y_1+y_2-1)}(y_2-y_1)}\right] \bigg\} \nonumber \\
& + \{y_1 \leftrightarrow y_2\}\ .
\end{align}
Note, by taking the $Q\to 0$ limit in (\ref{gen}), we can readily reproduce the amplitude of quarkonium photoproduction \cite{photoNLO1}.

At the high energy limit, the leading contribution to the NLO correction comes from the region $\eta\ll |x|\ll 1$, and the amplitude can be simplified to
\begin{align}
\mathcal{M}^{++}=\mathcal{M}^{--}&\approx-\frac{2m}{Q}\mathcal{M}^{00}\approx \dfrac{4i\pi^2\sqrt{4\pi\alpha}e_q m}{3\eta(m^2+\tfrac{Q^2}{4})}\left(\dfrac{\langle O\rangle_V}{m}\right)^{1/2} \Bigg[\alpha_sF^g(\eta,\eta,t)\nonumber \\
&+ \dfrac{\alpha_s^2}{\pi}\text{ln}\dfrac{m^2+\tfrac{Q^2}{4}}{\mu_F^2}\left(3\int_\eta^1\dfrac{dx}{x}F^g(x,\eta,t)
+\frac{4}{3}\int_\eta^1dxF^{q(+)}(x,\eta,t)\right)\Bigg].
\label{hlimit}
\end{align}
This expression suggests a suitable value of factorization scale, $\mu_F=\sqrt{m^2+\tfrac{Q^2}{4}}$. By taking this value, the NLO correction in (\ref{hlimit}) vanishes.

\section{Numerical analysis}

For full numerical evaluation, the knowledge of GPD over the full range of $x$ (or at least $x>\eta$) is required.
Unfortunately, the available models for GPD are fraught with uncertainties, especially in the ERBL region.
In order to minimize the uncertainties, we take the Forward Model (FM) in the DGLAP region, based on which
the imaginary sector of amplitude can be calculated. The FM tells
\begin{align}
&H^g(x,\eta,\mu_0)=xg(x, \mu_0),\quad \text{for}\ x>\eta\ ,\nonumber \\
&H^q(x, \eta, \mu_0)=q(x, \mu_0),\quad \text{for}\ x>\eta\ ,
\label{ansatz}
\end{align}
with $\mu_0=1$ GeV as initial scale, and MSTW08 \cite{mstw} as input PDF.
The GPDs at the energy scale of our concern are obtained through skewed evolution equation, where the NLO evolution kernels employed come from \cite{evo}. From the imaginary part of the amplitude, real part can be restored via the derivative dispersion relation (DDR) \cite{dispersion}:
\begin{equation}
{\rm Re}\frac{\mathcal{M}}{s}\approx {\rm tan}\left(\frac{\pi}{2}\frac{d}
{d{\rm ln}s}\right){\rm Im}\frac{\mathcal{M}}{s}
\approx \frac{\pi}{2}\frac{d}{d{\rm ln}s}{\rm Im}\frac{\mathcal{M}}{s}\ .
\label{dispersion}
\end{equation}

In numerical evaluation, those contributions from $E^p$, $\tilde{E}^p$ and $\tilde{H}^p$ are neglected, due to the following reasons. First, these terms are kinematic or helicity suppressed. Secondly, a rough numerical estimation tells that those terms contribute less than $1\%$ of the total. Moreover, there still lack convinced models to evaluating these GPDs.

Other parameters taken in the calculation go as follows:
\begin{itemize}
\item $\Lambda_{\text{QCD}}^3=332$ MeV\ ,\ $\Lambda_{\text{QCD}}^4=292$ MeV\ ,\ $\Lambda_{\text{QCD}}^5=210$ MeV\ ;
\item $|R_{J/\psi}(0)|^2=0.903$ GeV$^3$\ ,\ $|R_{\Upsilon}(0)|^2=7.76$ GeV$^3$\ ;
\item $1.4$ GeV$\leq m_c\leq1.6$ GeV\ ,\ $4.8$ GeV$\leq m_b\leq5.0$ GeV\ ;
\item $\text{max}\big[\frac{1}{2}\sqrt{m^2+\frac{Q^2}{4}}, \mu_0\big] \leq \mu_F \leq 2 \sqrt{m^2+\frac{Q^2}{4}}$\ ;
\item $\mu_R=\mu_F$\ .
\end{itemize}

The renormalization scale in our evaluation is set to be equal to the factorization scale, by which contributions from terms proportional to $\beta_0 \ln\frac{\mu_R}{\mu_F}$ are eliminated (in $c_0$s of Eqs. (\ref{appeq1}) and (\ref{appeq2}) in the Appendix).
The theoretical uncertainties are estimated by varying the values of quark mass and factorization scale.

\begin{figure}
\centering
\subfigure[]{
\includegraphics[width=0.45\textwidth]{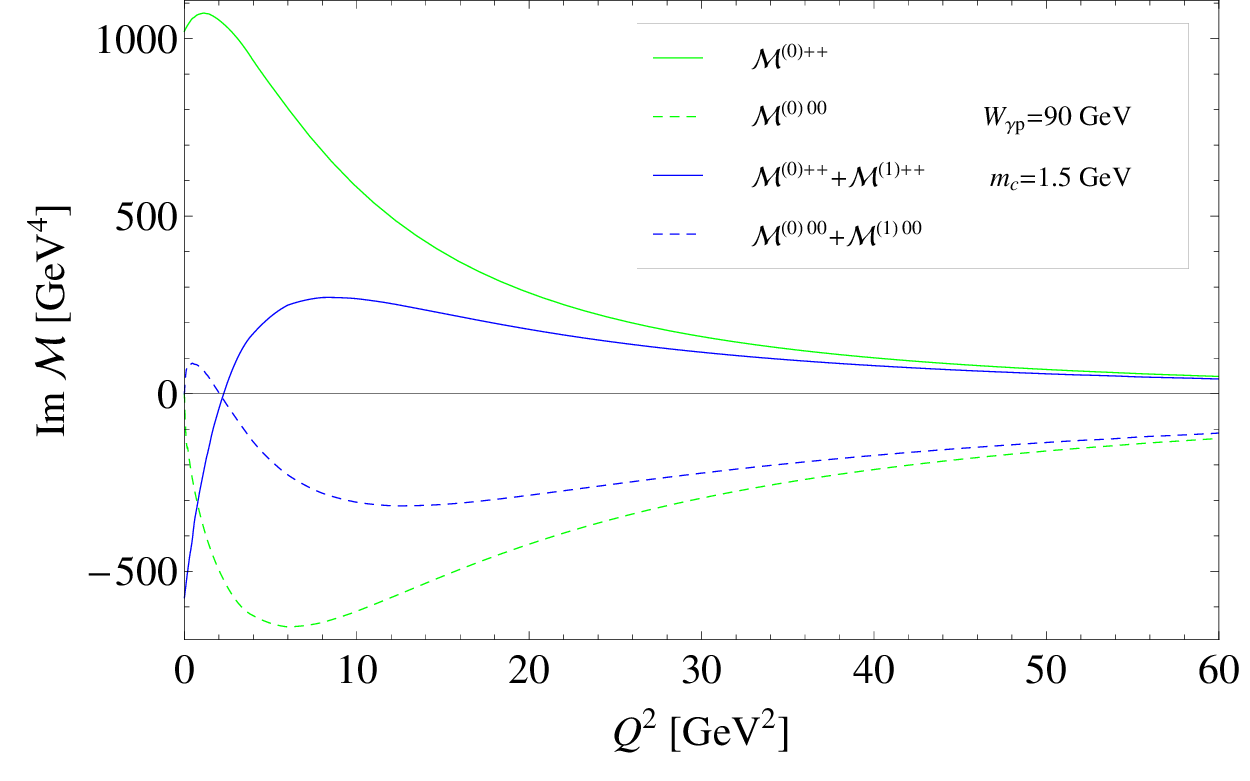}}
\subfigure[]{
\includegraphics[width=0.45\textwidth]{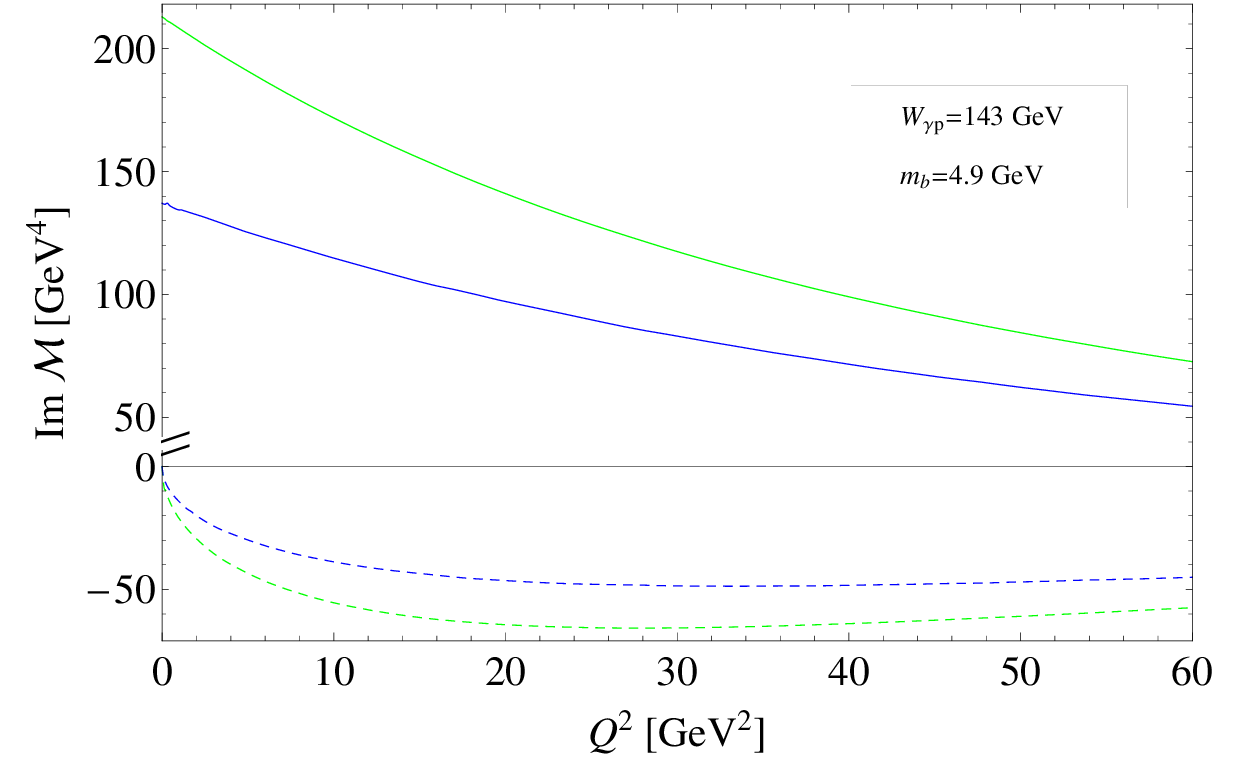}}
\caption{The imaginary parts of amplitudes of $J/\psi$ electroproduction (a) and $\Upsilon$ electroproduction (b). The factorization scale $\mu_F=\sqrt{m^2+\tfrac{Q^2}{4}}$. The LO and LO+NLO amplitudes are represented by green and blue lines respectively, of which the solid lines for transverse polarization and dashed lines for longitudinal polarization.}
\label{Im}
\end{figure}

The imaginary parts of the amplitudes in $J/\psi$ electroproduction as function of $Q^2$ are shown in Figure \ref{Im}(a), where the perturbative convergence exhibits. As $Q^2$ increases, the perturbative convergence is improved due to the increase of energy scale and the departure of small $x_B$ region. At large $Q^2$, say $Q^2 > 10$ GeV$^2$, the convergence manifests itself well. At low $Q^2$, the convergence is poor, especially when $Q^2 < 2$ GeV$^2$, where the full LO+NLO amplitude has opposite sign to the LO one. In the region $Q^2\approx 2$ GeV$^2$, the absolute value of LO+NLO amplitudes are very small, and hence the cross section as well. For $\Upsilon$ production, due to the large quark mass, the convergence works well even for the photoproduction, as demonstrated in Figure \ref{Im}(b).

\begin{figure}
\centering
\subfigure[]{
\includegraphics[width=0.45\textwidth]{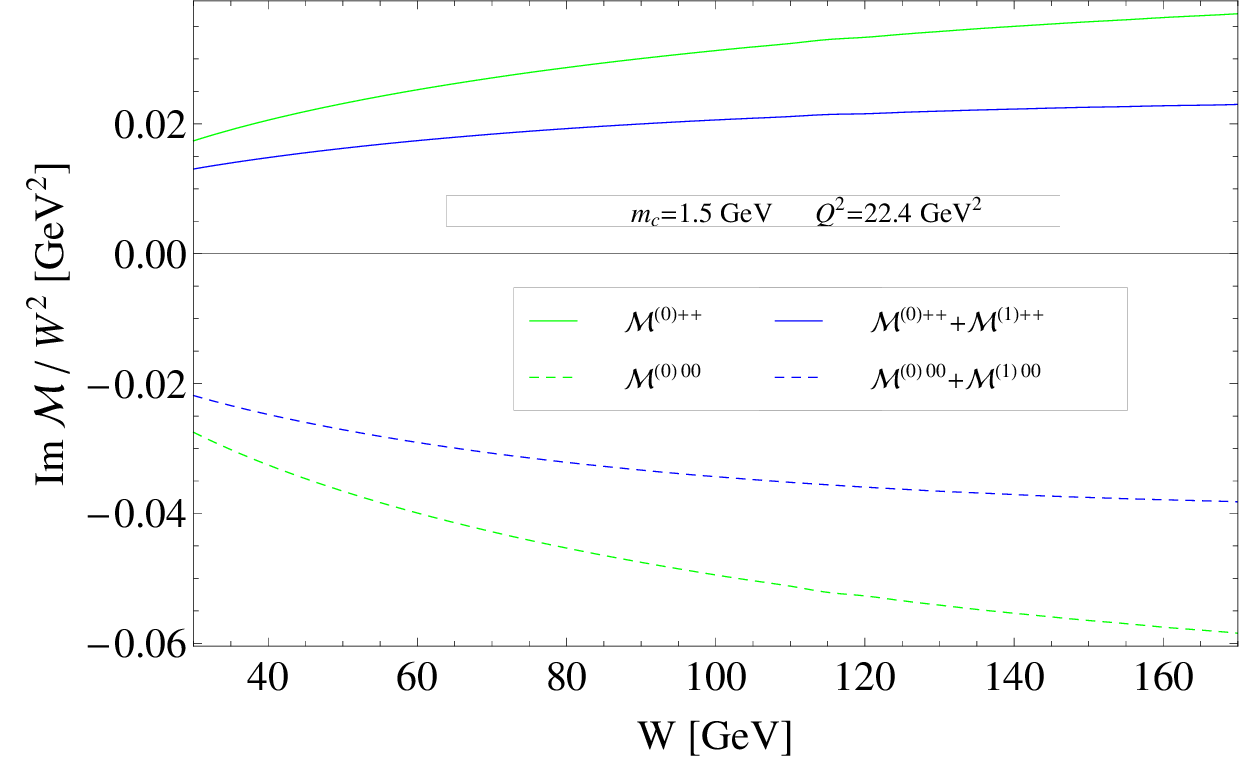}}
\subfigure[]{
\includegraphics[width=0.45\textwidth]{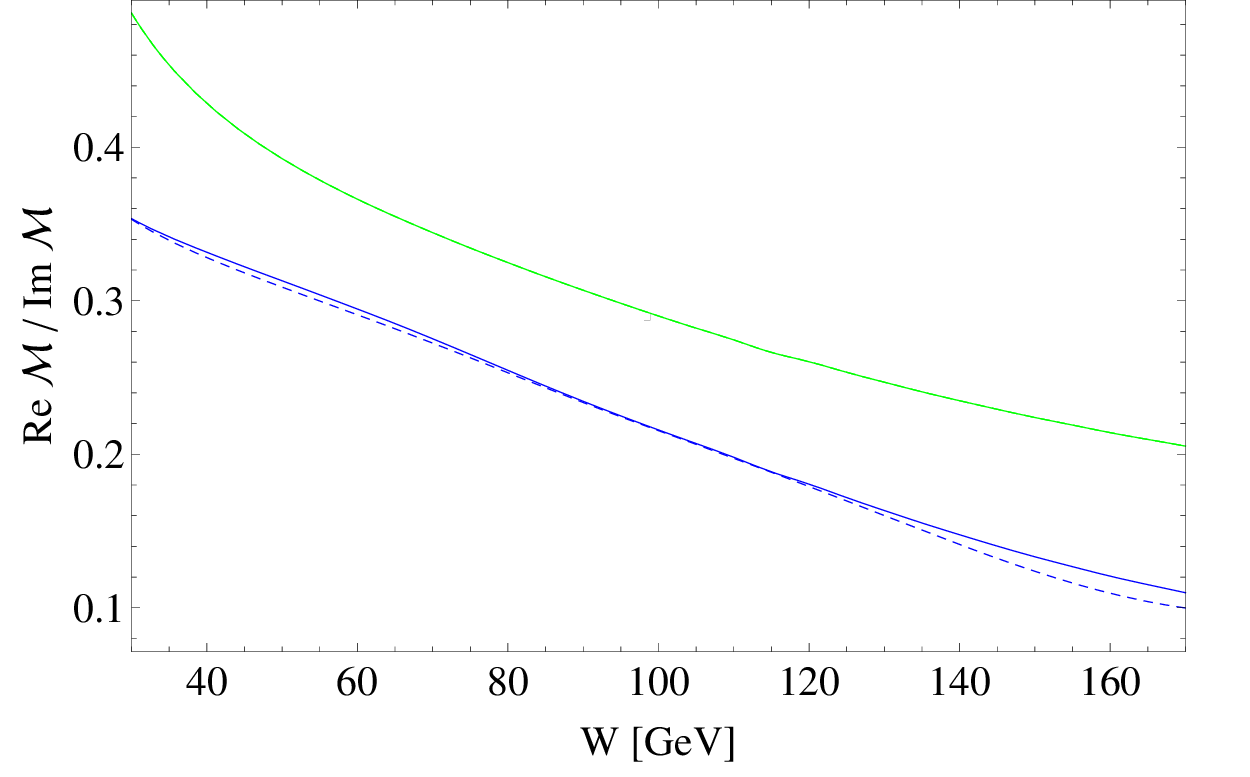}}
\caption{Predictions of  ${\rm Im}\mathcal{M}/W^2$ (a) and ${\rm Re}\mathcal{M}/{\rm Im}\mathcal{M}$ (b) for $J/\psi$ electroproduction at $Q^2=22.4$ GeV$^2$ as function of $W$. Here, the factorization scale $\mu_F=\sqrt{m^2+\tfrac{Q^2}{4}}$. The LO and LO+NLO amplitudes are represented by green and blue lines respectively, of which the solid lines denote for transverse polarization and dashed lines for longitudinal polarization. Note, the solid-green and dashed-green lines in (b) overlap since $\mathcal{M}^{(0)00}=-\tfrac{Q}{M}\mathcal{M}^{(0)++}$.}
\label{ReIm}
\end{figure}

Predictions of ${\rm Im}\mathcal{M}/W^2$ for $J/\psi$ electroproduction at $Q^2=22.4$ GeV$^2$ as function of $W$ are shown in Figure \ref{ReIm} (a), based on which we can estimate the error of DDR (\ref{dispersion}).
The DDR was derived from the standard integral dispersion relation (IDR) \cite{IDR1,IDR2} by taking the $s\to \infty$ limit and neglecting the subtraction constant term. An extended derivative dispersion relation (EDDR) given in Ref. \cite{EDDR} is equivalent to IDR.
In practice, when $s$ is big enough but finite the applicability and accuracy of (\ref{dispersion}) depend upon the behavior of ${\rm Im}\mathcal{M}/W^2$. By fitting ${\rm Im}\mathcal{M}/W^2$ to the form of $a_1W^{b_1}+a_2W^{b_2}$ and calculating ${\rm Re}\mathcal{M}$ via the DDR and EDDR separately, we find the discrepancy in amplitudes of these two methods is less than $1\%$.
Besides, in fact the variable $s$ in (\ref{dispersion}) should be replaced by $\nu =s+\frac{Q^2}{2}-\frac{M^2}{2}$,
and the corresponding corrections are then suppressed by a factor of $\tfrac{Q^2-M^2}{2s}$, which is less than $10^{-2}$ in our case.
The ratios of real parts, restored via (\ref{dispersion}), to imaginary parts as function of $W$ are shown in Figure \ref{ReIm} (b).

\begin{figure}
\begin{center}
\includegraphics[width=0.56\textwidth]{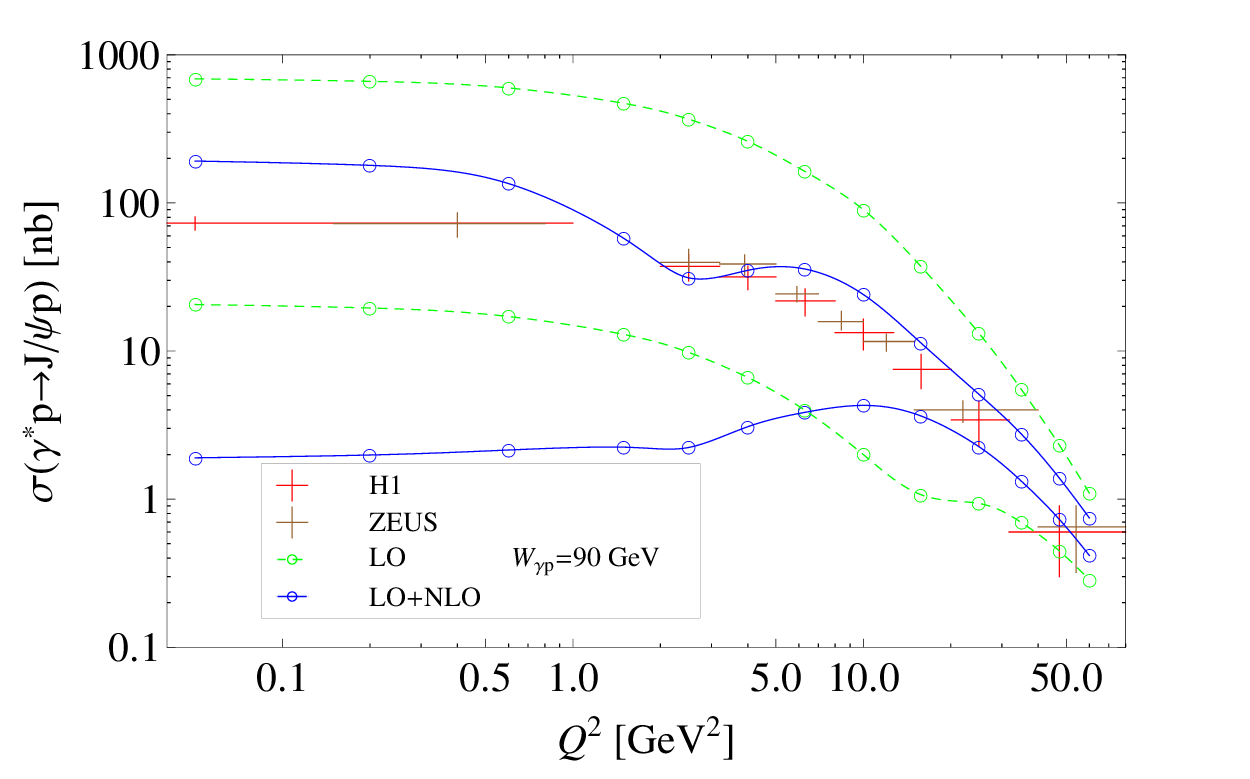}
\caption{The total cross section of exclusive $J/\psi$ electroproduction as function of $Q^2$ at $W_{\gamma p}=90$ GeV. The experimental data come from H1 \cite{H1} and ZEUS \cite{ZEUS} measurements, represented by red error bar and brown error bar respectively. The LO and up-to-NLO theoretical predictions are represented by double-dashed-green and double-solid-blue lines, referring to the upper and lower bounds of uncertainties, respectively.}
\label{JW90}
\end{center}
\end{figure}

\begin{figure}
\centering
\subfigure[]{
\includegraphics[width=0.3\textwidth]{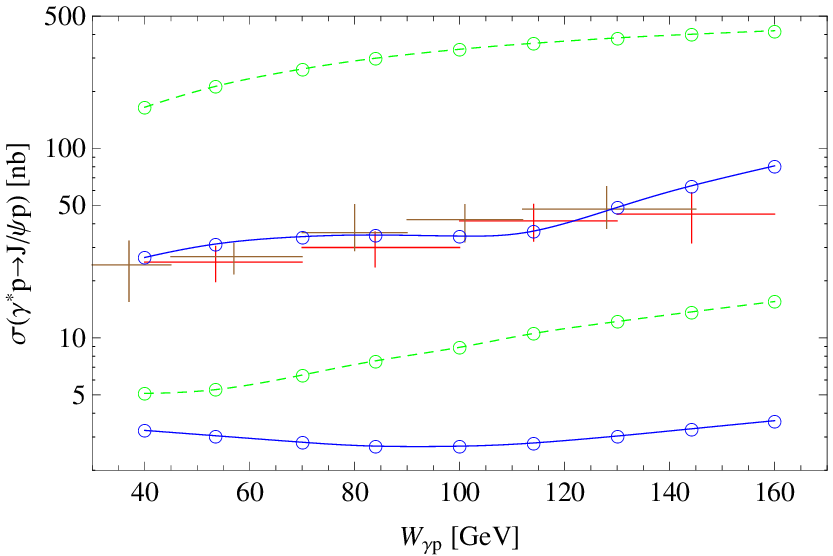}}
\subfigure[]{
\includegraphics[width=0.3\textwidth]{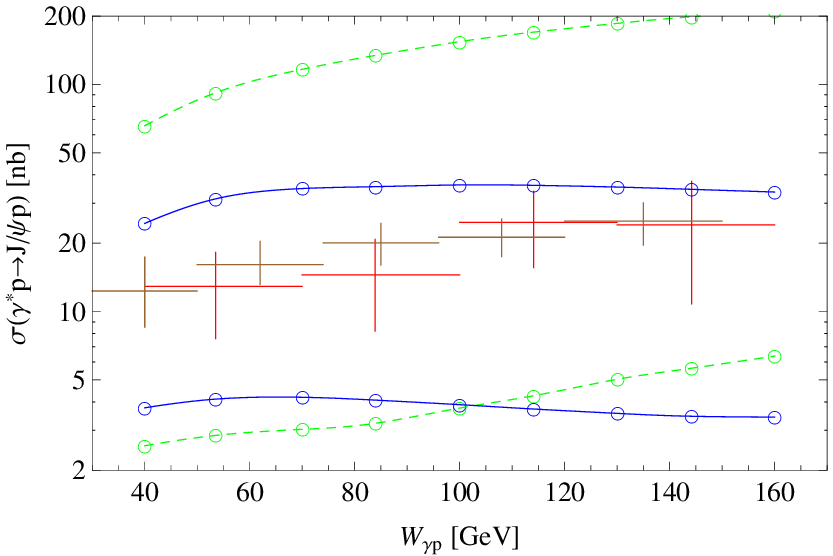}}
\subfigure[]{
\includegraphics[width=0.3\textwidth]{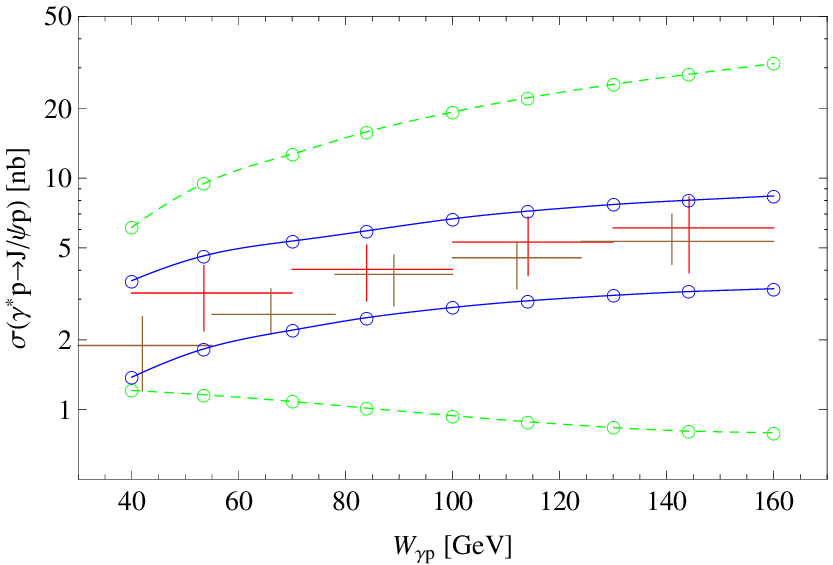}}
\caption{The total cross section of exclusive $J/\psi$ electroproduction as function of $W$ with different $Q^2$. (a) $Q^2 = 3.2\ \text{GeV}^2$, (b) $Q^2 = 7.0\ \text{GeV}^2$, (c) $Q^2 = 22.4\ \text{GeV}^2$. Notations for different lines and error bars are the same as Figure \ref{JW90}.}
\label{JQ123}
\end{figure}

Confronting to the HERA experimental condition, we calculate the differential cross section $d\sigma/dt$ at $|t|=|t|_{\text{min}}$, i.e. $\Delta_\perp=0$. Taking $t$ dependence measurement from H1 experiment as input \cite{H1}, we obtain the total cross section with different $W$ and $Q$. The numerical results of $J/\psi$ electroproduction are shown in Figure \ref{JW90} and \ref{JQ123}.
The $\psi(2S)$ feed down contribution is not taken into account in our calculation, since in experiment \cite{H1,ZEUS} this effect was subtracted in the data. Although the LO prediction may somehow cover the experimental data, the uncertainties are dubiously large, which mainly comes from the strong dependence of GPD on the factorization scale $\mu_F$, especially at small $\eta$. With NLO\footnote{Strictly speaking, matrix element squared $|\mathcal{M}^{(0)}+\mathcal{M}^{(1)}|^2$ includes some of the NNLO contributions.} corrections, at $Q^2 > 10$ GeV$^2$, the uncertainty is greatly suppressed as expected, and the prediction agrees with the experimental measurement well. At low $Q^2$, there are not many experimental data and the theoretical predicability is impaired by the large uncertainty, which remains even with NLO corrections.
Agreeing with above amplitude convergence analysis, numerical results for cross section in the region of $Q^2 < 5$ GeV$^2$ are unreliable, which casts a shadow on the perturbative QCD evaluation of $J/\psi$ photoproduction. Note, the dips on blue lines near $Q^2 \approx 2$ GeV$^2$ in Figure \ref{JW90} are understandable because the amplitude with NLO corrections drops off at this point, as shown in Figure \ref{Im}.

\begin{figure}
\begin{center}
\includegraphics[width=0.56\textwidth]{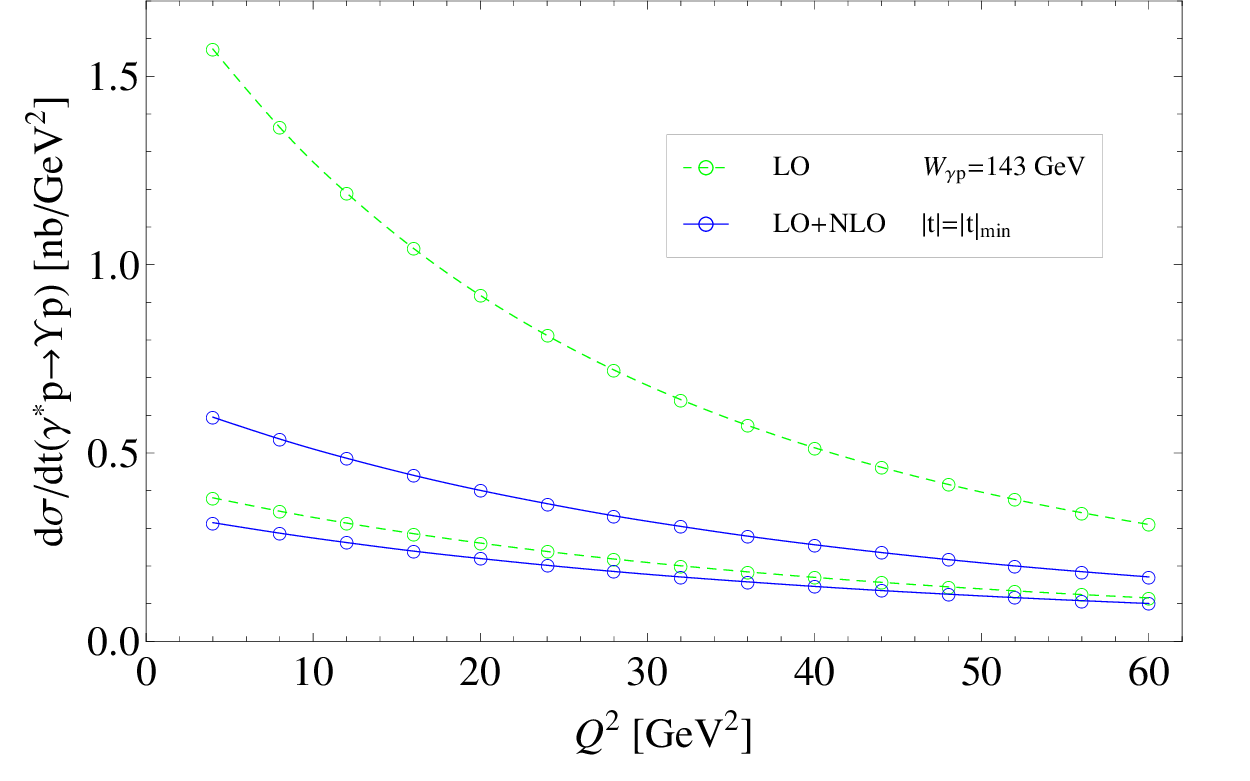}
\caption{Differential cross section $d\sigma/dt$ of exclusive $\Upsilon$ electroproduction as function of $Q^2$ at $|t|=|t|_{\text{min}}$ and $W=143$ GeV.}
\label{JW143}
\end{center}
\end{figure}

In comparison to the $J/\psi$ production, the theoretical evaluation of $\Upsilon$ production will definitely be more reliable because of the higher energy scale herein. However, unfortunately, to date there still has been no experimental result on the $\Upsilon$ leptoproduction yet. But hopefully it would be investigated on future lepton-nucleon colliders. For this aim, we calculate the $Q^2$ dependence of $d\sigma/dt$ at $|t|=|t|_{\text{min}}$ for $\Upsilon$ production in HERA experimental condition for illustration, as shown in Figure \ref{JW143}.

\section{Summary and conclusions}

We calculated analytically the exclusive electroproduction of quarkonium in the NRQCD framework and collinear factorization scheme up to the NLO QCD accuracy. In order to compare with experimental measurements, numerical evaluation about the cross section of exclusive $J/\psi$ electroproduction at different $Q^2$ and $W$ were performed. We estimated the theoretical uncertainties by varying the magnitudes of heavy quark mass and factorization scale.

At large $Q^2$, say $Q^2>10$ GeV$^2$, the NLO corrections may greatly reduce the theoretical uncertainty, and hence enable the predictions more reliable. We found a good agreement with the H1 \cite{H1} and ZEUS \cite{ZEUS} data. At small $Q^2$, say $Q^2< 5$ GeV$^2$, the pQCD analysis on exclusive $J/\psi$ production tends to be dubious. To make a prediction for future experiment on $ep$ collision, we schematically calculated the exclusive $\Upsilon$ electroproduction in HERA condition. In this case the quark mass is much heavier and guarantees the legitimacy of pQCD use even in the photoproduction.

To understand more about the parton distribution in nucleon, the GPD here, is one of the motivations of our study. With reliable theoretical calculation, people expect to have some feedbacks on GPD by confronting to the experimental measurement.
About the GPD, in the calculation we made use of the Forward Model (\ref{ansatz}) together with the NLO GPD evolution equation to evaluate GPD at DGLAP region. The input PDF and the initial scale were set to be MSTW08 \cite{mstw} at $\mu_0=1$ GeV. As a trial, another set of input PDF, the CT14 \cite{ct14} and initial scale $\mu_0=1.3$ GeV were applied. We found at reasonably large $\eta$ and $\mu_F$ region, say $\mu_F>2.4$ GeV and $\eta>0.001$, the difference of the cross sections with those two sets of input PDFs and initial scales is less than $15\%$, which does not influence our conclusions on $J/\psi$ electroproduction at large-$Q^2$ and $\Upsilon$ electroproduction. The main reason of this is because that the discrepancy between MSTW08 and CT14 diminishes with the increase of energy scale in small $x_B$ region. We therefore conclude that the Forward Model is simpler, parameter-free and adequate to explain the data.

Last, as an one step further investigation, we also calculated the concerned processes by using the GPD model proposed by Freund, McDermott, and Strikeman (FMS) \cite{gpdmodel}, the Shuvaev transform \cite{shuvaev1,shuvaev2} approach and the Double Distribution (DD) model \cite{dd1,dd2,dd3}, and found:

1) The FMS model agrees with the Forward model (\ref{ansatz}) in the DGLAP region, while in the ERBL region it is simply an ansatz based on the polynomiality of the lowest Mellin moments. In this model, the real part of the amplitude can be directly calculated without employing the dispersion relation. For large-$Q^2$ $J/\psi$ electroproduction and $\Upsilon$ electroproduction, the FMS model may yield a similar numerical result as the Forward Model does with about $10\%$ of discrepancy.

2) The Shuvaev transform can correlate the GPD to an auxiliary function, the usual parton distribution function in our evaluation.
By performing the transform, the skewed effect from evolution (at least LO evolution\footnote{Hopefully also the higher order evolution in the conformal scheme ($\overline{\rm CS}$).}) would be incorporated as well. The Shuvaev transform can also be viewed approximately as the LO GPD evolution of the $\eta$-dependent initial distribution. However, with the progress of evolution, the discrepancy between the $\eta$-dependent initial distribution and the initial distribution in Forward model shrinks, while the discrepancy may be enlarged by different evolution equations. Numerically, except for the $x\to 1$ region where the parton distributions tend to be diminished, the discrepancy between GPD from Shuvaev-transform approach and what from the Forward model is less than $15\%$ at moderate scale ($\mu>2$ GeV), and the difference in amplitude is about $8\%$ and $30\%$ at LO and NLO respectively.

3) The DD model is also a widely used GPD model. It provides an natural way to implement the polynomiality, nevertheless is unstable in evolution. It may yield a huge result stemming from the enhanced quark contribution, which should be tamed somehow.

\newpage
\noindent
{\bf Acknowledgements}

This work was supported in part by the Ministry of Science and Technology of the Peoples' Republic of China(2015CB856703); by the Strategic Priority Research Program of the Chinese Academy of Sciences, Grant No.XDB23030100; and by the National Natural Science Foundation of China(NSFC) under the Grants 11375200 and 11635009.

\newpage
\section*{Appendix}
The nonzero coefficients $c_i$ in (\ref{gen}).

For $A_g^{++}$($A_g^{--}$),\ $c^{++}_i=c^{--}_i=c_i$, with
\begin{align}
&c_0=\tfrac{1}{1728}\tfrac{1} {(y_1+y_2)^2}\tfrac{1}{2y_1+2y_2-1}
(-\tfrac{26 \pi ^2 y_1^3}{y_2^2}+\tfrac{192 y_1^3}{2 y_2-1}+
\tfrac{72 \pi ^2 y_1^3} {y_2}+\tfrac{35 \pi ^2 y_1^2}{y_2^2}-
\tfrac{194 \pi ^2 y_1^2}{y_2}+608 \pi ^2 y_1^2 \nonumber \\
&\quad +3264 y_1^2-\tfrac{11 \pi ^2 y_1}{y_2^2}+3168 y_1 y_2-
\tfrac{48 y_1}{2 y_2-1}+536 \pi ^2 y_1 y_2-\tfrac{54 \pi ^2 y_1}{y_2}-260
\pi ^2 y_1-1584 y_1 \nonumber \\
&\quad +\tfrac{71 \pi ^2}{y_2}-177 \pi ^2)-\tfrac{1}{y_1+y_2}
\tfrac{\beta_0}{24}\text{ln}\tfrac{\mu_R^2}{\mu_F^2}\ ,
\nonumber \\
&c_1=\tfrac{1}{36}(\tfrac{8 y_1}{y_2}-\tfrac{1}{y_1 y_2}-
\tfrac{1}{y_1-1}-\tfrac{2}{y_1}-\tfrac{38}{y_2}+8),\quad c_2=
\tfrac{1}{72}(\tfrac{1}{y_1 y_2}+\tfrac{29}{(y_1-1) y_2}+
\tfrac{2}{y_1}+\tfrac{6}{y_1-1}+\tfrac{22}{y_2})\ ,
\nonumber \\
&c_3=\tfrac{1}{36}\tfrac{y_1+y_2-1}{(y_1-y_2)^2(y_1+y_2)^2}
(-\tfrac{8 y_1^4}{y_2}+\tfrac{33 y_1^3}{y_2}-8 y_1^3+16 y_1^2 y_2+
\tfrac{16 y_1^2}{y_2}-60 y_1^2-5 y_1 y_2+16 y_1)\ ,
\nonumber \\
&c_4=\tfrac{1}{144}\tfrac{y_1+y_2-1}
{(y_1+y_2)^3}(-\tfrac{5 y_1^2}{y_2^2}-\tfrac{44 y_1^2}{y_2}+
\tfrac{15 y_1}{y_2^2}-\tfrac{80 y_1}{y_2}-4 y_1-\tfrac{11}{y_2}-47)\ ,
\nonumber \\
&c_5=\tfrac{1}{72}\tfrac{1}{(y_1-y_2)^{2}}\tfrac{1}{y_1+y_2}
(-\tfrac{8 y_1^4}{y_2}+\tfrac{106 y_1^3}{y_2}+16 y_1^2 y_2-
\tfrac{38 y_1^2}{y_2}-111 y_1^2+\tfrac{6 y_2^3}{2 y_1-1}-\tfrac{4 y_2^3}
{(2 y_1-1)^2}+\tfrac{2 y_2^3}{y_1} \nonumber \\
&\quad -\tfrac{21 y_2^2}{2 y_1-1}+\tfrac{6 y_2^2}{(2 y_1-1)^2}+
\tfrac{y_2^2}{y_1}-51 y_1 y_2+\tfrac{y_1}{2 y_2}+\tfrac{27 y_2}{2 (2 y_1-1)}-
\tfrac{3 y_2}{(2 y_1-1)^2}+\tfrac{1}{4 (2 y_1-1) y_2}+\tfrac{79 y_1}{2} \nonumber \\
&\quad-\tfrac{11}{4 (2 y_1-1)}+\tfrac{1}{2 (2 y_1-1)^2}-
8 y_2^3+83 y_2^2-\tfrac{49 y_2}{2}+\tfrac{1}{4 y_2}-\tfrac{13}{4})\ ,
\nonumber \\
&c_6=\tfrac{1}{72}\tfrac{1}{(y_1-y_2)^{2}}\tfrac{1}{y_1+y_2}\tfrac{1}
{(2y_1+2y_2-1)^{2}}(\tfrac{28 y_1^5}{y_2}+\tfrac{24 y_1^5}{2 y_2-1}-
\tfrac{16 y_1^5}{(2 y_2-1)^2}-\tfrac{84 y_1^4}{y_2}-\tfrac{100 y_1^4}{2 y_2-1}+
\tfrac{24 y_1^4}{(2 y_2-1)^2}\nonumber \\
&\quad+264 y_1^4+216 y_1^3 y_2+\tfrac{63 y_1^3}{y_2}+\tfrac{78 y_1^3}{2 y_2-1}-
\tfrac{12 y_1^3}{(2 y_2-1)^2}-462 y_1^3 -8 y_1^2 y_2^2-174 y_1^2 y_2-
\tfrac{14 y_1^2}{y_2}\nonumber \\
&\quad-\tfrac{23 y_1^2}{2 y_2-1}+\tfrac{2 y_1^2}{(2 y_2-1)^2}+
195 y_1^2+11 y_1 y_2+\tfrac{3 y_1}{2 y_2-1}-15 y_1)\ ,
\nonumber \\
&c_7=\tfrac{y_1}{2y_2(y_1+y_2)},\quad c_8=\tfrac{-y_1^2-y_2^2}{2y_2(y_1+y_2)^2}\ ,
\nonumber \\
&c_9=\tfrac{1}{288}\tfrac{1}{(y_1+y_2)^{2}}(-\tfrac{5 y_1^2}{y_2^2}-
\tfrac{9 y_2^2}{y_1^2}-\tfrac{40 y_1^2}{y_2}+\tfrac{13 y_2}{y_1^2}+
\tfrac{15 y_1}{y_2^2}+\tfrac{12 y_2^2}{y_1}-\tfrac{94 y_1}{y_2}-
\tfrac{38 y_2}{y_1}+252 y_1+\tfrac{33}{y_1}\nonumber \\
&\quad +16 y_2+\tfrac{35}{y_2}-118)\ ,
\nonumber \\
&c_{10}=\tfrac{1}{144}\tfrac{1}{(y_1+y_2)^2}(\tfrac{20 y_1^2}{y_2}+
\tfrac{15 y_1}{y_2}-4 y_1-\tfrac{24}{y_2}+13)\ ,
\nonumber \\
&c_{11}=\tfrac{1}{144}\tfrac{(y_1-5y_2)(y_2-5y_1)}{(y_1-y_2)^2(y_1+y_2)^2}
(\tfrac{y_1^3}{y_2^2}-\tfrac{3 y_1^2}{y_2^2}+\tfrac{15 y_1^2}{y_2}-
24 y_1^2+8 y_1 y_2-\tfrac{4 y_1}{y_2}+8 y_1-1)\ .
\label{appeq1}
\end{align}

For $A_g^{00}$, $c^{00}_i=-\sqrt{y_1+y_2-1}c_i$, with
\begin{align}
&c_0=\tfrac{1}{216}\tfrac{1}{(y_1+y_2)^2}(-\tfrac{14 \pi ^2 y_1}{y_2^2}-
\tfrac{24 y_1}{2 y_2-1}+\tfrac{22 \pi ^2 y_1}{y_2}-\tfrac{12}{(2 y_1-1) (2 y_2-1)}+16 \pi ^2 y_1+168 y_1-
\tfrac{24}{2 y_2-1}\nonumber \\
&\quad -\tfrac{37 \pi ^2}{y_2}+42 \pi ^2-12)-\tfrac{1}{y_1+y_2}
\tfrac{\beta_0}{24}\text{ln}\tfrac{\mu_R^2}{\mu_F^2}\ ,
\nonumber \\
&c_1=\tfrac{-2y_1-7}{9y_1y_2},\quad c_2=\tfrac{4 y_1^2+14 y_1-7}{18 (y_1-1) y_1 y_2}\ ,
\nonumber \\
&c_3=\tfrac{1}{9}\tfrac{1}{(y_1-y_2)^2}\tfrac{1}{(y_1+y_2)^2}
(\tfrac{2 y_1^4}{y_2}+\tfrac{5 y_1^3}{y_2}-3 y_1^3+17 y_1^2 y_2-
\tfrac{7 y_1^2}{y_2}-14 y_1^2-23 y_1 y_2+23 y_1)\ ,
\nonumber \\
&c_4=\tfrac{1}{9}\tfrac{y_1+y_2-1}{(y_1+y_2)^3}
(-\tfrac{2 y_1^2}{y_2}-\tfrac{7 y_1}{y_2}-2 y_1+\tfrac{1}{y_2}-3)\ ,
\nonumber \\
&c_5=\tfrac{1}{18}\tfrac{1}{(y_1-y_2)^2}\tfrac{1}{y_1+y_2}
(\tfrac{16 y_1^3}{y_2}+\tfrac{2 y_1^2}{y_2}-\tfrac{49 y_1^2}{2}-
\tfrac{5 y_2^2}{2 (2 y_1-1)}+\tfrac{2 y_2^2}{(2 y_1-1)^2}+
\tfrac{7 y_2^2}{y_1}-9 y_1 y_2-\tfrac{y_1}{2 y_2}\nonumber \\
&\quad +\tfrac{5 y_2}{2 (2 y_1-1)}-\tfrac{y_2}{(2 y_1-1)^2} -
\tfrac{1}{4 (2 y_1-1) y_2}+\tfrac{47 y_1}{4}-\tfrac{21}{8 (2 y_1-1)}+
\tfrac{19 y_2^2}{2}-\tfrac{15 y_2}{2}-\tfrac{1}{4 y_2}-\tfrac{21}{8})\ ,
\nonumber \\
&c_6=\tfrac{1}{18}\tfrac{1}{(y_1-y_2)^2}\tfrac{1}{y_1+y_2}\tfrac{1}
{2y_1+2y_2-1}(\tfrac{14 y_1^3}{y_2}-\tfrac{5 y_1^3}{2 y_2-1}+
\tfrac{4 y_1^3}{(2 y_2-1)^2}-2 y_1^3-30 y_1^2 y_2-\tfrac{7 y_1^2}{y_2}+\tfrac{7 y_1^2}{2 y_2-1}\nonumber \\
&\quad-\tfrac{2 y_1^2}{(2 y_2-1)^2}+29 y_1^2+\tfrac{31 y_1 y_2}{2}-
\tfrac{25 y_1}{4 (2 y_2-1)}-\tfrac{65 y_1}{4}-\tfrac{1}{2 (2 y_2-1)}-\tfrac{1}{2})\ ,
\nonumber \\
&c_7=\tfrac{y_1}{2 y_2 (y_1+y_2)},\quad c_8=\tfrac{-y_1^2-y_2^2}{2 y_2 (y_1+y_2)^2}\ ,
\nonumber \\
&c_9=\tfrac{1}{36}\tfrac{1}{(y_1+y_2)^2}(-\tfrac{18 y_1^2}{y_2}-
\tfrac{7 y_2}{y_1^2}+\tfrac{12 y_1}{y_2}+\tfrac{20 y_2}{y_1}+4 y_1-
\tfrac{15}{y_1}-14 y_2-\tfrac{8}{y_2}+32)\ ,
\nonumber \\
&c_{10}=\tfrac{1}{36}\tfrac{1}{(y_1+y_2)^2}(\tfrac{18 y_1^2}{y_2}-
\tfrac{12 y_1}{y_2}+18 y_1+\tfrac{3}{y_2}-8)\ ,
\nonumber \\
&c_{11}=\tfrac{1}{18}\tfrac{(y_1-5y_2)(y_2-5y_1)}{(y_1-y_2)^2(y_1+y_2)^2}
(-y_1^2+3 y_1 y_2+\tfrac{y_1}{y_2}-4 y_1+1)\ .
\label{appeq2}
\end{align}

For $A_q^{++}$($A_q^{--}$), $c^{++}_i=c^{--}_i=c_i$, with
\begin{align}
&c_0=\tfrac{\pi ^2}{9 y_2 (y_1+y_2)^2},\ c_3=-\tfrac{8 (y_1+y_2-1)}
{9 (y_1-y_2) (y_1+y_2)^2},\  c_4=\tfrac{2 (y_1+y_2-1)}{3 y_2
(y_1+y_2)^3},\  c_5=\tfrac{2 (y_1^2+4 y_1 y_2-y_1-y_2^2-y_2)}{9 (2 y_1-1) y_2 (y_1-y_2) (y_1+y_2)}
,\nonumber \\
& c_6=-\tfrac{2 (2 y_1^2-18 y_1 y_2+7 y_1+5 y_2-2)}{9 (2 y_2-1)
(y_1-y_2) (y_1+y_2) (2 y_1+2 y_2-1)},\quad c_7=-\tfrac{4}{9 y_2
(y_1+y_2)},\quad c_8=c_9=-\tfrac{4 (y_2-y_1)}{9 y_2 (y_1+y_2)^2}
\ ,\nonumber \\
&c_{10}=-\tfrac{4 y_1-3}{9 y_2 (y_1+y_2)^2},\quad c_{11}=
-\tfrac{2 y_1 (12 y_1 y_2-3 y_1-4 y_2^2-y_2)}{9 y_2 (y_1-y_2) (y_1+y_2)^2}\ .
\end{align}

For $A_q^{00}$, $c^{00}_i=-\sqrt{y_1+y_2-1}c_i$, with
\begin{align}
&c_3=\tfrac{8 (y_1+y_2-2)}{9 (y_1-y_2) (y_1+y_2)^2},\quad c_5=
-\tfrac{4 (2 y_1^2-y_1-y_2)}{9 (2 y_1-1) y_2 (y_1-y_2) (y_1+y_2)},
\quad c_6=-\tfrac{8 (y_2-1)}{9 (2 y_2-1) (y_1-y_2) (y_1+y_2)}
\ ,\nonumber \\
&c_7=c_{10}=-\tfrac{4}{9 y_2 (y_1+y_2)},\quad c_8=c_9=
-\tfrac{4 (y_2-y_1)}{9 y_2 (y_1+y_2)^2},\quad c_{11}=
-\tfrac{8 y_1 (y_1-3 y_2+2)}{9 (y_1-y_2) (y_1+y_2)^2}.
\end{align}

For $\tilde{A}_g$, $\tilde{c}^{00}_i=0,\tilde{c}^{++}_i=-\tilde{c}^{--}_i=c_i$, with
\begin{align}
&c_0=-\tfrac{(10 \pi ^2 y_2+24 y_2-5 \pi ^2)(y_2-y_1)}{432 y_2 (2 y_2-1) (y_1+y_2)}
,\quad
c_1=\tfrac{1}{36 (y_1-1)}
,\quad
c_2=-\tfrac{5}{36 (y_1-1)}
,\quad
c_4=-\tfrac{5 (y_1+y_2-1)}{36 y_2 (y_1+y_2)}
\ ,\nonumber \\
&c_5=\tfrac{(38 y_1^2-6 y_1 y_2-21 y_1+5 y_2)}{36 (2 y_1-1)^2 (y_1+y_2)}
,\quad
c_6=\tfrac{(y_1-y_2) (12 y_1 y_2-10 y_1-22 y_2+13)}{36 (2 y_1-1) (2 y_2-1)^2 (y_1+y_2)}
,\quad
c_9=\tfrac{(3 y_2-5 y_1)}{72 y_2 (y_1+y_2)}
\ ,\nonumber \\
&c_{11}=-\tfrac{(y_1-5 y_2) (y_1-y_2) (5 y_1-y_2)}{72 y_1 y_2 (y_1+y_2)}\ .
\end{align}

For $\tilde{A}_q$,  $\tilde{c}^{00}_i=0,\tilde{c}^{++}_i=-\tilde{c}^{--}_i=c_i$, with
\begin{align}
c_5=-\tfrac{2 (y_1-y_2)}{9 (2 y_1-1) y_2 (y_1+y_2)}
,\quad
c_6=\tfrac{4 (y_1+y_2-1)^2}{9 (2 y_1-1) (2 y_2-1) (y_1+y_2) (2 y_1+2 y_2-1)}\ .
\end{align}
\end{document}